\documentstyle[twocolumn,aps,prl]{revtex}
\begin{document}
\draft
\title{Energy Gap Structure in Bilayer Oxide Superconductors}
\author{Kazuhiro Kuboki* and Patrick A. Lee}
\address{Department of Physics, Massachusetts Institute of Technology\\
Cambridge, Massachusetts, 02139\\}

\maketitle

\begin{abstract}
We consider a model for bilayer superconductors where an interlayer
pairing
amplitude $\Delta_\perp$ co-exists with intralayer pairing
$\Delta_\|$ of $d_{x^2 - y^2}$ symmetry.  This model is motivated by a
recent
photoemission experiment reporting the splitting of the nodes of the
energy
gap.  In addition to offering a natural explanation of this observation,
the model has a number of new experimental consequences.  We find that
the
new state is accompanied by a spontaneous breaking of the tetragonal
symmetry.  We also find that the out-of-phase oscillation of
$\Delta_\perp$ and $\Delta_\|$ gives rise to a new Raman active mode.
The phase of $\Delta_\perp$ may also become imaginary, leading to a
state
which breaks time reversal symmetry, which may have important
implications
for tunnelling experiments.
\end{abstract}

In the past two years, there has been mounting evidence for nodes in the
gap in oxide superconductors \cite{Hardy,Imai}.  Recently, a new class
of
experiments which are phase sensitive have provided strong evidence for
a
sign change of the order parameter as a function of
angle \cite{Wollman,Tsuei}, consistent with $d$ symmetry.
 At the same time, there are several experiments which appear
to be inconsistent with the simple $d_{x^2-y^2}$ state.  For example,
Sun et al \cite{Sun}, have made tunnel junctions between
$YBa_2Cu_3O_{7-\delta}\  (YBCO)$ and $Pb$ and obtained Josephson current
along the
$c-$axis.  Secondly, a recent report of fractional vortices
tied to grain boundary interfaces \cite{Kirtley} has been interpreted
as requiring a time reversal symmetry breaking state near the
interface \cite{Sigrist1}. Finally, Ding et al \cite{Ding} measured
 the energy gap as a function of angle usng angular resolved
photoemission in
$Bi_2Sr_2CaCu_2O_8\ (Bi$-2212) and claim that the energy gap does
not vanish along
$(\pi, \pi)$ as expected for the $d_{x^2 - y^2}$ state.
Instead, the node is split
into two, lying approximately $10^\circ$ from the $45^\circ$ line.
Another group\cite{Ma} also reported a finite energy gap in the
$(\pi, \pi)$ direction and studied its temperature dependence.
  Taken together, these experiments
suggest that the energy gap structure may be more complicated than
$d_{x^2-y^2}$.  In this paper we show that a modification of the
$d_{x^2-y^2}$ state due to interlayer pairing provides a scenario
whereby
the experiments mentioned above can be explained.  Our picture also
leads
to a number of predictions which can be tested experimentally.

Band structure calculations have yielded a surprisingly large interlayer
hopping term between the bilayers, given by
$t_\perp (\vec k) c^\dagger_{1\sigma} (\vec k) c_{2\sigma} (\vec k)$
where 1 and 2 refer to the layers.  Chakravarty et al\cite{Chakravarty}
have
proposed the parametrization
$t_\perp (\vec k) = {1\over 4} t_\perp^\circ (\cos k_x a -\cos k_y a)^2$
with
$t_\perp^\circ \approx 0.24\  eV$.  Andersen et al \cite{Andersen}
recently showed that the unusual $\vec k$ dependence originates from
hopping
via the $Cu\ 4s$ orbitals, and that it leads to an interlayer exchange
energy of order $J_\perp \approx 20\ meV$, consistent with the lower
bound
of $8\ meV$ given by neutron scattering\cite{Shamoto}.  It has been
proposed\cite{Ubbens,Ioffe} that the relatively large exchange term,
when
enhanced by intralayer antiferromagnetic correlations, is responsible
for
the spin gap phase found in bilayer systems \cite{Millis}.  Furthermore,
Ubbens and Lee\cite{Ubbens} have argued that in the superconducting
phase,
 a pairing amplitude
$\Delta_{12} = \langle c_{i\uparrow} (\vec r) c_{2\downarrow} (\vec
r)\rangle$
appears in addition to the intraplane order parameter
$\Delta_{ii} (\vec\eta ) = \langle c_{i\uparrow} (\vec r)
c_{i\downarrow}
(\vec r + \vec\eta )\rangle , \ i = 1, 2$ which is of $d$ symmetry.
This leads to a quasi-particle dispersion relation
$\tilde E_\pm (\vec k) = (\xi^2_k + |\Delta_\| \pm \Delta_\perp
|^2)^{1\over 2}$
where
$\xi_k = \epsilon_k - \mu, \Delta_\|(\hat k)$ and $\Delta_\perp$
are proportional to
$\Delta_{11} = \Delta_{22}$ and $\Delta_{12}$ respectively.
Since the nodes of this state are given by the
zeros of $|\Delta_\| \pm \Delta_\perp |$,  if $\Delta_\|$ has
$d_{x^2 - y^2}$ symmetry, the nodes are shifted from the $45^\circ$
direction
and split into two nodes\cite{Ubbens}.
  This provides a natural explanation of the
observation  by Ding et al \cite{Ding}.

Encouraged by the experiment, we decided to re-examine the interlayer
pairing
model.  Ubbens and Lee\cite{Ubbens} included $J_\perp$ but not the
$t_\perp$ term in their consideration.  In this paper we add both the
$t_\perp$ and $J_\perp$ terms to the standard $t-J$ model.  Wheras
Ubbens
and Lee attempted to justify the appearance of $\Delta_{12}$
microscopically,
in this paper we take a more phenomenological approach and assume the
co-existence of $\Delta_\perp$ and $\Delta_\|$.  This is motivated by
experiment:  as far as we know, the present scenario is the only one
which
is consistent with both the node splitting\cite{Ding} and the sign
change
of the order parameter as $\hat k$ varies from $0 {\rm\ to\ } \pi/2$, as
required by the corner SQUID experiment\cite{Wollman}.  The purpose of
this
work is to explore the consequences of the assumed interlayer pairing
amplitude,
so that further experiments can confirm or falsify this picture.

We begin by treating the bilayer $t-J$ model (including $t_\perp$ and
$J_\perp$) using the slave boson mean field method, which is equivalent
to
the Gutzwiller approximation.
The tight binding band
$\epsilon (k) = - 2\tilde t (\cos k_x a + \cos k_y a))$ is split into
bonding and anti-bonding bands
$g_\pm = (c_1 \pm c_2)/\sqrt 2$
with dispersion
$\epsilon_\pm (\vec k) = \epsilon (\vec k) \pm \tilde t_\perp (\vec k)$,
where $\tilde t$ is of order $J$ and
$\tilde t_\perp (\vec k) \simeq X_0t_\perp (\vec k)$
where $X_0$ is of order $x$,
the doping concentration.  Thus, in the mean field theory the effective
interlayer hopping is reduced by $x$, simply because in the strongly
correlated metals, the electron, on average, must find a vacancy to hop.

	The resulting Fermi surface is shown in Fig 1 for $x = 0.15$.
As emphasized by Anderson \cite{Anderson}, in the normal state, coherent
hopping is really not possible between the layers.  We expect the
bonding-antibonding splitting to be
smeared out, but a region of low lying excitations may exist in the
$\vec k$ space between the two Fermi surfaces.  Fig. 1 bears a striking
resemblance to the photoemission data of Dessau et al\cite{Dessau1}.

In the superconducting state, coherent hopping between the planes indeed
occurs, and we should take the band splitting seriously. With the basis
sets
$(c_{1, \vec k\uparrow}, c^\dagger_{1, -\vec k\downarrow},
c_{2, \vec k\uparrow}, c_{2, -\vec k\downarrow}^\dagger)$,
the mean field Hamiltonian takes the form
\begin{equation}
\left [\begin{array}{clcr}
\epsilon (\vec k) - \mu & \Delta_\| (\hat k) &
\tilde t_\perp (\vec k ) & \Delta_\perp \\
\Delta_\|^* (\hat k) & -\epsilon (\vec k) + \mu &
\Delta_\perp^* & -\tilde t_\perp (\vec k) \\
\tilde t_\perp (\vec k) & \Delta_\perp &
\epsilon (\vec k) - \mu & \Delta_\| (\hat k) \\
\Delta_\perp^* & -\tilde t_\perp (\vec k) &
\Delta_\|^* (\hat k) & -\epsilon (\vec k) + \mu
\end{array}\right ]
\end{equation}
It is easily seen that the bonding anti-bonding bands $g_\pm$ block
diagonalize this matrix, so that the $g_\pm$ bands are separately paired
by
$\Delta_\pm (\hat k) = \Delta_\| (\hat k) \pm\Delta_\perp$, resulting
in the quasi-particle spectrum
\begin{equation}
E_\pm (\vec k) = (\xi^2_\pm (\vec k) + |\Delta_\|
(\hat k) \pm \Delta_\perp |)^{1\over 2}
\end{equation}
where $\xi_\pm (\vec k) = \epsilon_\pm (\vec k) - \mu$.
We can choose $\Delta_\|$ to be
real and positive.  If $\Delta_\perp$ is also real and positive, the
nodes are
split as before, but the split nodes are associated with the $\pm$ bands
separately, as indicated in Fig. 1.  It is clear from Fig. 1 that for
real
$\Delta_\perp$, the onset of interlayer pairing implies a spontaneous
breaking
of the tetragonal symmetry of the model. This is a consequence of the
inclusion of $t_\perp$. The two degenerate states of the
broken symmetry correspond to $\Delta_\perp$ being positive or negative.
We shall refer to these degenerate states as $d \pm s$.  Due
to the vanishing of $t_\perp (\vec k)$ along $(\pi, \pi )$, this
assymmetry
is difficult to resolve near the nodes.  However, in a given domain, the
$g_+$ band has a different gap along $MY$ than $\overline M Y$ and the
resulting
asymmetry in the electronic state is expected to couple linearly to the
orthorhomic strain
$\epsilon = (a - b)/ (a + b)$.

There is one additional complication to this discussion, in that we
should
consider the possibility that $\Delta_\perp$ is purely imaginary.  We
shall
refer to this state as $d + is$, which
 has a minimum gap of $|\Delta_\perp |$.  Indeed, a mean field
calculation shows that this state is lower in energy than $d + s$.
The Ginzburg-Landau free energy is given by $F = F_0 + F_\epsilon$,
where
for simplicity we have fixed
$\Delta_{11} = \Delta_{22} = \Delta_\|$,
\begin{eqnarray}
F_0 = a_\| |\Delta_\| |^2 + b_\| |\Delta_\| |^4 + a_\perp |\Delta_\perp
|^2 +
b_\perp
 |\Delta_\perp |^4\nonumber\\
+\  d_1 |\Delta_\| |^2 |\Delta_\perp |^2 +
d_2 (\Delta^2_\| \Delta^{2*}_\perp + c.c.)
\end{eqnarray}

\begin{equation}
F_\epsilon = \alpha \epsilon (\Delta_\| \Delta^*_\perp + c.c.)
\end{equation}
where $F_\epsilon$ describes the linear coupling to the strain discussed
earlier.\cite{Sigristueda}
  In mean field theory, we found that $d_1 = 4 d_2 > 0$.
If we write $\Delta_\perp = |\Delta_\perp | e^{i\phi}$, the $d_2$ term
is
proportional to $\cos 2\phi$ with a positive coefficient, and is
minimized by
$\phi = {\pi\over 2}$.  The $F_\epsilon$ term, on the other hand, is
proportional to $\cos\phi$, so that $\phi = 0$ or $\pi$
(the  $d \pm s$ state) is stabilized in systems with pre-existing
orthorhomic
distortions such as $YBCO$.  In tetragonal systems, we add an elastic
energy
term $F_E = {1\over 2} \kappa\epsilon^2$ to $F$ and minimize with
respect
 to $\epsilon$.
This produces a term
$-2 (\alpha^2/\kappa ) |\Delta_\perp |^2 |\Delta_\| |^2 \cos^2\phi$
which opposes the $d_2$ term. The $d + s$ state is stabilized provided
${\alpha^2\over\kappa} > 2d_2$.
The observation of
split nodes in the nominally tetragonal $Bi$-2122 presumably means that
this
condition is satisfied and $d + s$ is stabilized.
Thus we predict that, for bilayer systems with tetragonal symmetry,
a superconducting state with split nodes is accompanied by a
spontaneous breaking of the tetragonal
symmetry.
The $Bi - 2122$ compound is not truly tetragonal due to the superlattice
modulation in the $(\pi, \pi)$ direction, but the $a$ and $b$ lattice
constants are predicted to become unequal at low temperature.
  The best test of the prediction
is probably in tetragonal materials such as the bilayer mercury
compounds.
Even in $YBCO$, we expect an additional distortion below
$T_c$.  There is evidence for this in the literature\cite{You}.
The coupling to lattice distortions also lead us to expect anomalies
in the transverse ultra sound velocity below $T_c$\cite{Sigristueda}.

As mentioned earlier, the $d + s$ state explains
the photoemission data of ref. 8.  Indeed, the lower branch min
$(\Delta_+ , \Delta_-)$ shown in the insert of Fig. 1
can be fitted to the experimental
data with $\Delta_\| (\hat k = 0) \approx 30 \ meV$ and
$\Delta_\perp \approx 5 {\rm \ to \ } 9 \ meV$ \cite{Norman}.
  In ref. 8, only data associated with the
$g_-$ band (dashed line in Fig. 1) was shown.
  The question arises as to what happened to the
upper branch of the insert in Fig. 1.
Two possibilities need to be examined:  the photoemission may be from a
single
domain of $d + s$ state or from multiple domains of $d + s$ and $d - s$.
In the
first case we expect a single peak with different gaps
$(\Delta_\| (\hat k = 0) \pm \Delta_\perp )$ at the $M - Y$ and
$\overline M - Y$ crossings.
In the second case the peak should be a superposition
of two peaks with a splitting of $2\Delta_\perp$.  The data is
consistent with
a splitting of $10 \ meV$ or less \cite{Norman} which may require a
slight angular
dependence of $\Delta_\perp (\hat k )$ so that it is smaller along
$\Gamma - M$ than along $\Gamma - Y$.
 At present, data is not available which covers both
$\Gamma - M$ and $\Gamma - \overline M$ quadrants in the same sample, so
that
these two possibilities cannot be distinguished.

The existence of several order parameters should lead to new collective
modes, which are amplitude and phase modes associated with
$\Delta_{ij} = |\Delta_{ij}| e^{i\phi_{ij}}, i, j = 1, 2$.  The
amplitude
modes are expected to be high in frequency and damped and we shall focus
on the phase modes only.  There are three modes corresponding to the
three phase degrees of freedom.

(1) The in-phase oscillation of
$\Delta_{11}, \Delta_{22}, \Delta_{12}$ is the
Bogoliubov-Anderson mode which is coupled to total charge density and
pushed up to the plasma frequency.

(2) The out-of-phase mode
$\phi_{11} - \phi_{22}$.
Recently this mode was discussed\cite{Wu} in the context of the
interlayer
tunnelling model of Chakravarty et al\cite{Chakravarty}.  However, as
pointed
out by Wu and Griffin, the existence of this mode is a general feature
of two layers coupled by Josephson tunnelling and was described by
Leggett\cite{Leggett}, who noted that the usual phase-number
commutation relation
$[\phi_{ii}, N_i ] = 2i$
implies a coupling of
$\phi_{11} - \phi_{22}$ to $N_1 - N_2$,
{\em i.e.}, to charge transfer between the planes.
The mode frequency $\omega_J$ can be computed by combining the
Josephson relations $\dot\phi = 2\ eV$ and
$\dot n_{12} = E_J \sin\phi$
with $eV = \gamma n_{12}$,
where $\gamma = C^{-1} + d\mu/dn$.
$n_{12}$ is the interlayer density oscillation,
$C$ is the capacitance per area, and
$E_J \approx |\tilde t_\perp |^2 N^2 (0) |\Delta_\| | a^2$
is the Josephson energy per area between the planes.  We find
$\omega_J = (2 E_J\gamma )^{1\over 2}$
which is just the Josephson plasma frequency.
If we ignore the capacitance term and approximate
$\gamma = d\mu/d n$ by $N (0)^{-1} \approx Ja^2$,
we find that
$\omega_J \approx xt_\perp (\Delta_\|/J)^{1\over 2}$
which is quite high, so that this mode is probably strongly damped.

(3) Finally, the mode which is of greatest interest to us is the
out-of-phase
oscillation between $\phi_{12}$ and $\phi_{11} = \phi_{22}$.
Since the free energy given by Eq. (3) depends weakly on this phase
difference, we expect this mode to be low-lying in frequency.
Furthermore,
this phase mode leads to a re-arrangement of the quasi-particle spectrum
in the plane, since a change in $\phi_{12}$ from $0$ to ${\pi\over 2}$
interpolates between the $d + s$ and $d + is$ states.  This should
couple
to charge oscillation within each plane, but with the total charge
conserved within each layer.  Thus, we expect this mode to be Raman
active.
To confirm this qualitative picture, we have carried out a collective
mode
calculation in a model where we introduce phenomenological attractive
coupling constants to stabilize $\Delta_\|$ and $\Delta_\perp$.  We
treat
this model in mean field theory and carry out an expansion around
the mean field in a standard way\cite{Wu}.  For simplicity we have set
$\epsilon = 0$.  We find that the globally stable states correspond to
either $\Delta_\|$ or $\Delta_\perp = 0$, while the
$d + is$ state lies slightly higher in energy but is locally stable.
The
global energy balance depends on details of the interaction and the
presence of $\epsilon$.  For example, a sufficiently large $\epsilon$
will stabilize the $d + s$ state because of the linear coupling in
Eq. (4).  Since our goal in the part of the analysis is to understand
the collective mode rather than the microscopic origin of the energy
gaps,
we proceed to expand about the locally stable $d + is$ state.  We find
that the Josephson mode
$\phi_{11} - \phi_{22}$
decouples as expected.  Setting
$\phi_{11} = \phi_{22}$, we find that
$\phi_{11}, \phi_{12}$ and $\rho_2$ are coupled, where $\rho_2$ is the
$l = 2$ component of the density fluctuation in the plane.  The
$3 \times 3$ matrix can be diagonalized.  Details of this calculation
will be given elsewhere, but the main result is an estimate of the
out-of-phase mode frequency, which turns out to be of order
$|\Delta_\perp |$.  For $|\Delta_\perp | \ll |\Delta_\| |$,
its damping should be small.  Furthermore, its coupling to
$\rho_2$ implies that it is Raman active in the $B_{1g}$
symmetry\cite{Wu}.
The detection of this mode will be a strong confirmation of the
existence of $\Delta_\perp$.

Next we examine what effects the more complicated order parameter
structure
have for tunnelling experiments.  In ref. 4, tunneling is between oxide
superconductors across a grain boundary.  A reasonable tunnelling
Hamiltonian
may be of the form
$H_t = t_0\displaystyle\sum_{i = 1, 2} c^\dagger_{iL} c_{iR} + c.c.$,
which will lead to a Josephson coupling of the form
\begin{equation}
H_J = T_J(\Delta^*_{+L} \Delta_{+R} + \Delta^*_{-L} \Delta_{-R} + c.c. )
\end{equation}

The two terms in Eq. 5 will generate two Josephson currents with
different dependences on
the grain boundary orientation \cite{Sigrist2}.
However, at present the half integer flux experiment in specially
designed
rings \cite{Tsuei} is done on twinned samples, so that the
$d \pm s$ domains are difficult to distinguish from the pure
$d_{x^2 - y^2}$ state.

In the experiment which measures the critical current of a
SQUID loop \cite{Wollman}, the
tunnel junctions are between the oxide superconductor and a conventional
superconductor, and the tunnelling current is in the $a - b$ plane.  In
this
case, it may be more reasonable to assume a tunnelling Hamiltonian of
the form
$H^\prime_T = t_0g^\dagger_+ c_R + c.c.$
where $c_R$ destroys an electron in the conventional superconductor.  In
this
case, only the symmetric order parameter $\Delta_+$ forms the Josephson
coupling.  Thus, in principle, if a single domain sample can be made
with
variable junction orientations, the shift of the node shown by the solid
line
in the insert of Fig. 1 can be detected.

We briefly discuss the experiment of ref. 5, where a Josephson current
is
observed along the $c$ axis between $YBCO$ and a conventional
superconductor.  The $d + s$ state by itself cannot explain why the
critical
current is relatively insensitive to whether the $YBCO$ is untwinned
or highly twinned.  This is because the coupling to the lattice strain
$\epsilon$ (shown in Eq. (4)) will lock the $d + s$ state to one set of
twins and the $d - s$ states to the other.  Since the $d$ order
parameter is
insensitive to twinning (otherwise the phase sensitive experiment of
references 3 and 4 will not work in twinned samples), the
Josephson current will have opposite signs on the two sets of twins and
tend
to cancel.  One possible way out of this dilemma is to postulate
the existence of regions where the
$d + is$ becomes stabilized.
  This is plausible due
to the small energy differences expected and can happen near the twin
boundaries.
  With this assumption a net Josephson current may be obtained.
We defer a detailed discussion of this
possibility to a later publication \cite{Kuboki1}.

The trapped fluxes on grain boundaries between $YBCO$ films with
different
orientations are measured and interpreted as being due to the appearance
of fractional charged vortices \cite{Kirtley}.   Such
vortices require the existence of a time-reversal symmetry breaking
state
near the interface \cite{Sigrist1}.  Recently, Kuboki and Sigrist
\cite{Kuboki2}
 offered an
explanation of this state as being due to an admixture with a proximity
induced $s$ component of the order parameter.  In our picture, the
possible
existence of the complex $d + is$ state near an interface suggests an
alternative origin of the time-reversal symmetry breaking state.  The
two
alternatives can be distinguished by searching for similar effects in
 single layer materials.

Finally, we address the issue of the possibility of a second phase
transition below $T_c$.  Usually, a Landau theory of the form given by
Eq. (3) with two order parameters implies two phase transitions, because
the
coefficients $a_\perp$ and $a_\|$ change signs at different
temperatures.
In the underdoped case, the transition is complicated by strong
fluctuation effects, so that, in some sense, the spin gap transition may
be considered the first transition, and the superconducting $T_c$ the
second transition.  This raises the interesting possibility that the
situation may be reversed in optimally doped or overdoped cases, which
would imply that the transition at $T_c$ is to a pure
$d_{x^2 - y^2}$ order paramter, followed at some lower temperature by
the
onset of interlayer pairing $(\Delta_\perp \ne 0)$.  Recently, a
second transition at $30K$ is reported in $T_2$ measurements in fully
oxygenated samples of $YBCO$ \cite{Itoh}.
Obviously, it will be interesting to look
for lattice distortions near this temperature.  Another interesting
possibility
is to study the fate of the fractional vortices at temperatures above
$30K$.

We thank Manfred Sigrist, Andy Millis and Mike Norman
 for many helpful discussions.  This
work was supported primarily by the Center for Materials Science and
Engineering under NSF grant DMR 94-00334.  K.K. acknowledges support by
the Ministry of Education, Science and Culture of Japan.

\figure{\ \\
\ \\
\ \\
\ \\
\ \\
\ \\
\ \\
\ \\
\ \\
\ \\
\ \\
\ \\
\ \\
\ \\
\ \\
\ \\
{\bf Figure 1.} Mean field calculation of the Fermi surface including
 hopping
 between bilayers for hole doping $x = 0.15$.  The parameters
$t/J = 3, t_\perp^{(0)}/J = 2.4$ and $J_\perp/J = 0.2$ are used.  The
solid
and dashed lines are the Fermi surfaces for the bonding $(g_+)$ and
antibonding
$(g_-)$ bands.  Solid and dashed arrows indicate the approximate
location of the nodes
associated with $\Delta_+$ and $\Delta_-$ in the $d + s$ state.
The insert shows a schematic picture of the angular dependence
($\theta = 0$ is along the $\Gamma - \overline M$ axis) of the gap
 functions
$|\Delta_+ |$ (solid) and
$|\Delta_-|$ (dashed) for
the $d + s$ state ($\Delta_\perp$ real and positive).
Note that
$\Delta_\pm$ are associated with the solid and dashed Fermi surfaces
respectively.  For the $d - s$ state, $|\Delta_+ |$ is given by the
dashed line.\label{Fig. 1}}

\end{document}